\documentclass[12pt,a4paper]{iopart}

\expandafter\let\csname equation*\endcsname\relax
\expandafter\let\csname endequation*\endcsname\relax
\usepackage{hyperref}

\usepackage{amsmath}    
\usepackage{amssymb}    
\usepackage{color}      
\usepackage{graphicx}   
\usepackage{bm}
\usepackage[sort&compress,numbers]{natbib}

\begin{document}

\title{Topology Regulation during Replication of the Kinetoplast DNA}

\author{%
Davide Michieletto,$^{1}$
Davide Marenduzzo,$^{2}$
and Matthew S. Turner$^{1}$
}

\address{%
$^{1}$ Department of Physics and Centre for Complexity Science, University of Warwick, Coventry CV4 7AL, United Kingdom.
$^{2}$ School of Physics and Astronomy, University of Edinburgh, Mayfield Road, Edinburgh EH9 3JZ, Scotland, United Kingdom.}

\begin{abstract}%
We study theoretically the replication of Kinetoplast DNA consisting of several thousands separate mini-circles found in organisms of the class Kinetoplastida. When the cell is not actively dividing these are topologically connected in a marginally linked network of rings with only one connected component. During cell division each mini-circle is removed from the network, duplicated and then re-attached, along with its progeny. We study this process under the hypothesis that there is a coupling between the topological state of the mini-circles and the expression of genetic information encoded on them, leading to the production of Topoisomerase. This model describes a self-regulating system capable of full replication that reproduces several previous experimental findings. We find that the fixed point of the system depends on a primary free parameter of the model: the ratio between the rate of removal of mini-circles from the network ($R$) and their (re)attachment rate ($A$). The final topological state is found to be that of a marginally linked network structure in which the fraction of mini-circles linked to the largest connected component approaches unity as $R/A$ decreases. Finally we discuss how this may suggest an evolutionary trade-off between the speed of replication and the accuracy with which a fully topologically linked state is produced. 

\end{abstract}

\maketitle

\section{Introduction}

The Kinetoplast DNA (or kDNA)~\cite{Fairlamb1978} is a network of linked DNA loops commonly found in a group of unicellular eukaryotic organisms of the class Kinetoplastida. Some of these organisms are responsible for human diseases such as sleeping sickness and leishmaniasis~\cite{Young1987,Jacobson2003,MacLean2004}. 
The Kinetoplast DNA (kDNA) is known for its unique structure, which resembles a percolating network of linked mini-circles at its critical point~\cite{Shapiro1995,Kellenberger1986,Chen1995,Michieletto2014a}. 

During replication, catenation between the loops introduces a non-trivial topological problem, which is solved as follows.
First Topoisomerase II disentangles one loop at a time from the network~\cite{Jensen2012}, the loop then diffuses through a region called the Kinetoflagellar Zone (KFZ), undergoes duplication via $\theta$ structure, and reaches the anti-podal sites, where, together with the progeny mini-circle, it links to the periphery of the network~\cite{Perez-Morga1993}. Both mini-circles are left with some nicks along their contour, so that they are no longer suitable for duplication~\cite{Perez-Morga1993}. This mechanism ensures that mini-circles are duplicated once and only once~\cite{Dna1984}. Finally, when the cell divides, two copies of the network are produced by slicing the network in two, through the middle~\cite{Perez-Morga1993,Liu2005}. This process has the unique feature that it must undergo complex topological changes in a precise and consistent order. Previous studies have shown that the action of Topoisomerases is crucial for this machinery to work correctly, not only for decatenation of mini-circles from the network, but also for the re-attachment at the end of duplication~\cite{Shlomai1983}. RNAi experiments showed that in absence of Topoisomerase, the Kinetoplast is unable to form and most of the mini-circles remain in a free (unlinked) state~\cite{Wang2000a}. Type II Topoisomerase is well-known to play a crucial role in simplifying knots and catenane which occur in DNA~\cite{White1987}. On the other hand, it has been shown that the same enzyme is also capable of creating complex, linked structures rather than just simplifying them~\cite{Hsieh1980,Kreuzer1980,Brown1981}. It has been conjectured that its role in this respect is related to DNA concentration and is mediated by the presence of polyamines, \textit{e.g.} spermidine~\cite{Krasnow1982}. 

In this work, we focus on the role of Topoisomerase in both adding and removing mini-circles to and from the network. We suggest that this process is intimately related to the local concentration of linked or free mini-circles together with the local concentration of Topoisomerase. The central idea of our work is that we model the latter as itself depending on the linkedness of the mini-circles via the local concentration of free mini-circles. This means that our model can describe the Kinetoplast replication as a self-regulating mechanism that, when switched on, can drive the removal, duplication and attachment of mini-circles without further intervention. Our model correctly reproduces key features observed during the Kinetoplast replication, as the centripetal movement of newly linked mini-circles toward the centre of the network and the asymmetric distribution of Topoisomerase which localises the edge of the network. Finally, we investigate the relationship between the  efficiency with which the Kinetoplast re-attaches the mini-circles and the speed at which the whole network duplicates. We argue that the balance between these two factors can shed some light on why the Kinetoplast requires some redundancy in its genome in order to survive.

\section{The Model}
\begin{figure*}[t]
\centering
\includegraphics[scale=0.35]{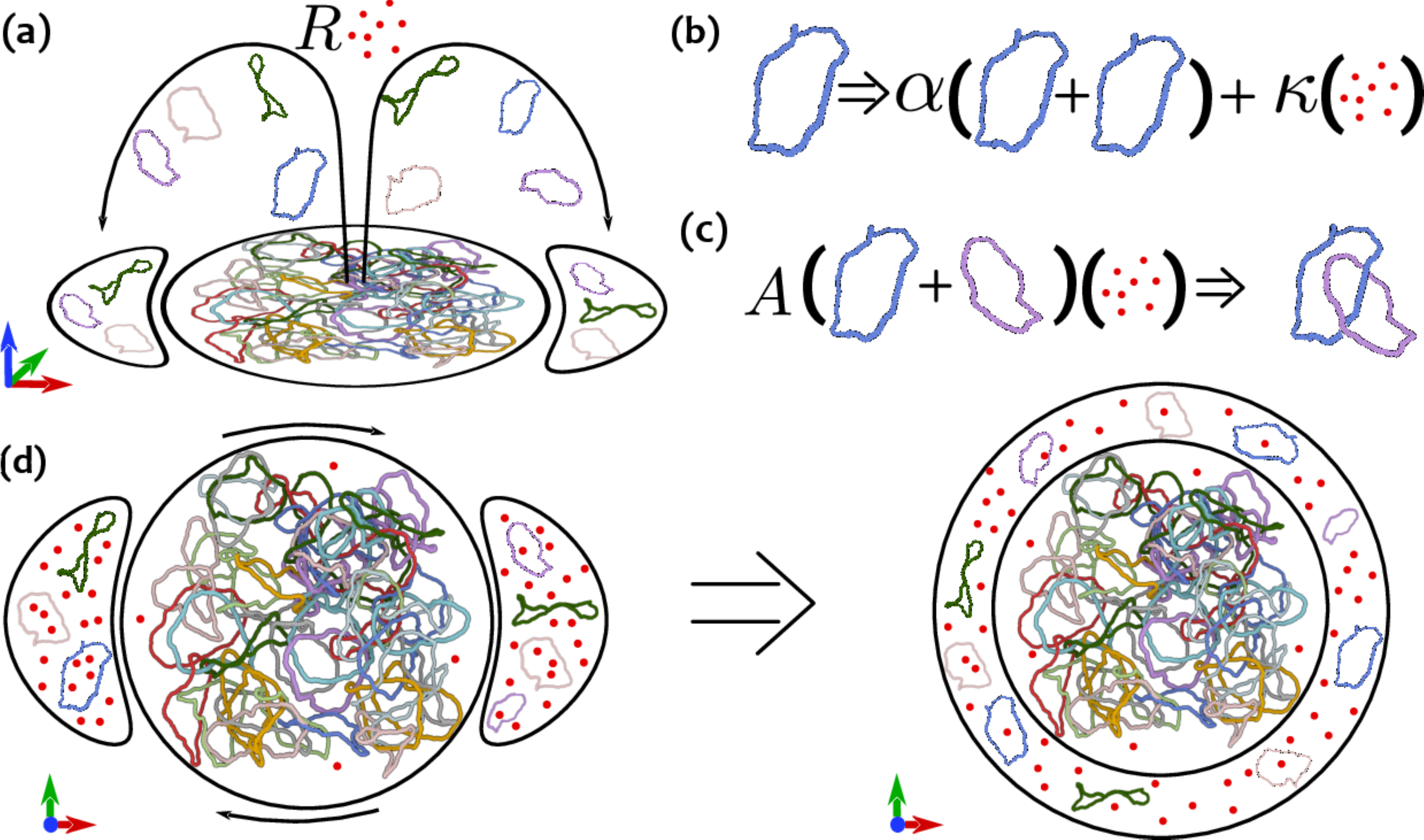}
\caption{\textbf{(a)} Graphical representation of the reaction in eq.~\eqref{eq:removal}. Topoisomerase (red dots) unlinks mini-circles (coloured loops) from the central connected network.  \textbf{(b)} Graphical representation of the reaction in eq.~\eqref{eq:duplication}. Each mini-circle duplicates at rate $\alpha$ and indirectly produces Topoisomerase (red dots) at rate $\kappa$. \textbf{(c)} Graphical representation of the reaction in eq.~\eqref{eq:reattachment}. Each mini-circle forms a catenane with another mini-circle (either free or linked) at a rate proportional to the local concentration of Topoisomerase. \textbf{(d)} We substitute the two anti-podal sites where duplication and re-attachment occurs with a shell. This can be done because of the relative rotation of the Kinetoplast core with respect to the anti-podal sites.  }
\label{fig:Panel1}
\end{figure*}

The replication of the Kinetoplast can be divided into three major steps, which are translated in our model as follows: (i) each mini-circle is unlinked from its neighbours in the network at a rate that is proportional to the \emph{local} concentration of Topoisomerase II (see Fig.~\ref{fig:Panel1}(a)):
\begin{equation}
(N) Linked \xrightarrow{R \phi_T} Unlinked + (N-1) Linked .
\label{eq:removal}
\end{equation}
(ii) While diffusing through the kinetoflagellar zone, each mini-circle is then duplicated at rate $\alpha$ and  transcribed thereby stimulating the downstream production of Topoisomerase II at rate $\kappa$, \emph{e.g.} via transcription factors encoded in the mini-circles~\cite{Wang2000a} (see Fig.~\ref{fig:Panel1}(b)):
\begin{align}
&Unlinked \xrightarrow{\alpha} Unlinked + Unlinked\notag\\
\text{ and } \notag \\
&Unlinked \xrightarrow{\kappa} Topoisomerase.
\label{eq:duplication}
\end{align}
(iii) Once a mini-circle reaches the anti-podal sites, it is then re-ligated and re-attached at the periphery of the network or linked together with another mini-circle at a rate again proportional to the \emph{local} concentration of linker enzyme, Topoisomerase II, (see Fig.~\ref{fig:Panel1}(c)): 
\begin{align}
&Unlinked + Unlinked \xrightarrow{A \phi_T} Linked \notag\\
\text{ or } \notag\\
&Unlinked + Linked \xrightarrow{A \phi_T} Linked.
\label{eq:reattachment}
\end{align}
The rates $R$ and $A$ are key parameters in our model and describe the rates of removal ($R$) of mini-circles from the network and their re-attachment ($A$). They will be further discussed in the following sections. Previous studies showed that during the replicating phase, the Kinetoplast rotates relatively to the anti-podal sites where re-attachment occurs. This has been thought to facilitate an even redistribution of progeny mini-circles around the periphery of the network~\cite{Perez-Morga1993,Jensen2012}. In our model, this feature is modelled by substituting the two anti-podal complexes with a shell which surrounds the Kinetoplast core (see Fig.~\ref{fig:Panel1}(d)). In fact, this is equivalent to performing a time-average of the relative position of the complexes with respect to the core over the whole replicating phase. In light of this, we can visualise the entire system as a rotationally symmetric 2-dimensional disk divided into a core $r\leq R_k$ filled by the Kinetoplast at the beginning of the replication ($t=0$) and a shell $R_k<r<R_{max}$ (initially empty) which plays the role of region where re-attachment occurs. \\

We describe the system in terms of the relative density of linked and unlinked mini-circles, \emph{i.e.} at the beginning of replication, the density of linked mini-circles is 1 ($\rho_l=1$) within the Kinetoplast, and there are no unlinked mini-circles ($\rho_u=0$). During the replicating phase, the number of mini-circles grows and we assume that, at the end of the replication, $\rho_l + \rho_u = 2$, which corresponds to a concentration of about $50$ $mg/ml$~\cite{Chen1995}. The equations describing our model in terms of these relative densities and in terms of the concentration of Topoisomerase $\phi_T$ are the following:

\begin{align}
\dfrac{d \rho_l}{dt} = &-R\rho_l \phi_T + A\phi_T\rho_u\left(\rho_l +\rho_u\right) + D_l \nabla^2 \rho_l \label{eq:Eqs.a}\\
\dfrac{d \rho_u}{dt} = &+R\rho_l \phi_T - A\phi_T\rho_u\left(\rho_l +\rho_u\right) + D_u \nabla^2 \rho_u + \notag\\
& + \alpha \rho_u \left[2 - \left(\rho_u+\rho_l\right)\right] \label{eq:Eqs.b} \\
\dfrac{d \phi_{T}}{dt} = &+\kappa \rho_u - \dfrac{\phi_{T}}{\tau} + D_{T} \nabla^2 \phi_{T}.
\label{eq:Eqs.c}
\end{align}\\

$D_l$, $D_u$ and $D_T$ are the diffusion coefficients of, respectively, linked and unlinked mini-circles and of Topoisomerase II enzymes. $R$ and $A$ are the rates of removal and attachment of mini-circles from and to the Kinetoplast. $\kappa$ and $\tau$ the rate of production and decay time of Topoisomerase II. Finally, $\alpha$ is the rate of duplication of mini-circles.\\

The last term in eq. \eqref{eq:Eqs.b} ensures that the mini-circles duplicate at rate $\alpha$ until the sum $\rho_u + \rho_l$ reaches $2$, at which point the growth is stopped. Biologically, this is ensured by the fact that re-attached mini-circles are left with gaps in order to prevent further duplication~\cite{Jensen2012}. The third terms in each equation describe the diffusion of linked and unlinked mini-circles and Topoisomerase II, respectively. Since unlinked mini-circles are freed from the network, we assume that they diffuse away from the Kinetoplast. Previous studies have reported that mini-circle duplication begins in the Kinetoflagellar Zone (KFZ), nearby the Kinetoplast, and it is always completed inside the anti-podal regions~\cite{Jensen2012}. The KFZ is composed by a dense matrix of filaments which attach the Kinetoplast outer membrane to the flagellar basal body~\cite{Jensen2012}. Because of this, we estimate the diffusion coefficient of the free mini-circles by computing the diffusion of a macro-molecule of size comparable to that of the mini-circles through a dense medium. The diffusion coefficient can be estimated with the Ogston formula~\cite{Ogston1973,Johnson1996}:
\begin{equation}
D_u = D_{\infty} \exp{\{-\phi^{1/2}r_{mc}/r_f\}},
\label{eq:Ogston}
\end{equation}
where $\phi$ is the volume fraction of filaments, $r_{mc}$ the gyration radius of mini-circles and $r_f$ the radius of the filaments. $D_\infty$ is the diffusion coefficient of the mini-circle in free solution and can be estimated via the Stokes-Einstein formula:
\begin{equation}
D_{\infty} = \dfrac{k_B T}{6 \pi \eta r_{mc}},
\end{equation}
where $\eta = 10$ $cP$ is typical the viscosity of the cytoplasm. By using $r_{mc}=\sqrt{L l_k/12}$ with $L \sim 2$ $kbp \sim 700$ $nm$  and $l_k\sim 100$ $nm$, at room temperature, one obtains $D_{\infty} \simeq 0.3$ $\mu m^2/s$. The KFZ structure has been found to be a rather dense region and it is thought to be responsible for keeping the Kinetoplast in place during replication and mitosis. Transmission electron microscopy has been performed to study the composition of the KFZ~\cite{Ogbadoyi2003}. This is highly filled with filaments, which radius is reported to be between 5 and 10 $nm$~\cite{Ogbadoyi2003}. This justifies the use of eq.~\eqref{eq:Ogston}, where we assume a modest volume fraction $\phi = 0.1$. This leads to a diffusion coefficient for the unlinked rings $D_u \simeq 0.001$ $\mu m^2/s$. On the other hand, mini-circles which are linked to the Kinetoplast are unable to freely diffuse, as they are topologically connected with thousands of other mini-circles. In light of this, we include in our model the diffusion of linked mini-circles with diffusion coefficients $D_l \simeq D_u/N_{mc}$, where $N_{mc}=5000$ is the number of mini-circles in the Kinetoplast. We also include the diffusion of Topoisomerase. We implicitly assume that as soon as Topoisomerase II is produced, it binds to either linked or unlinked rings nearby. In our model, the production of this enzyme is stimulated by the presence of unlinked rings (the only available for transcription) and therefore, it is highly likely that Topoisomerase binds to unlinked rings. When this happens, Topoisomerase executes one strand crossing, which in the case of unlinked rings, generates linked rings. This implies that in our model, Topoisomerase molecules are much less mobile than their free form, and they inherit their diffusion coefficient from the linked rings. In light of this we set $D_T \simeq D_l$. \\

\begin{figure*}
\centering
\includegraphics[scale=0.35]{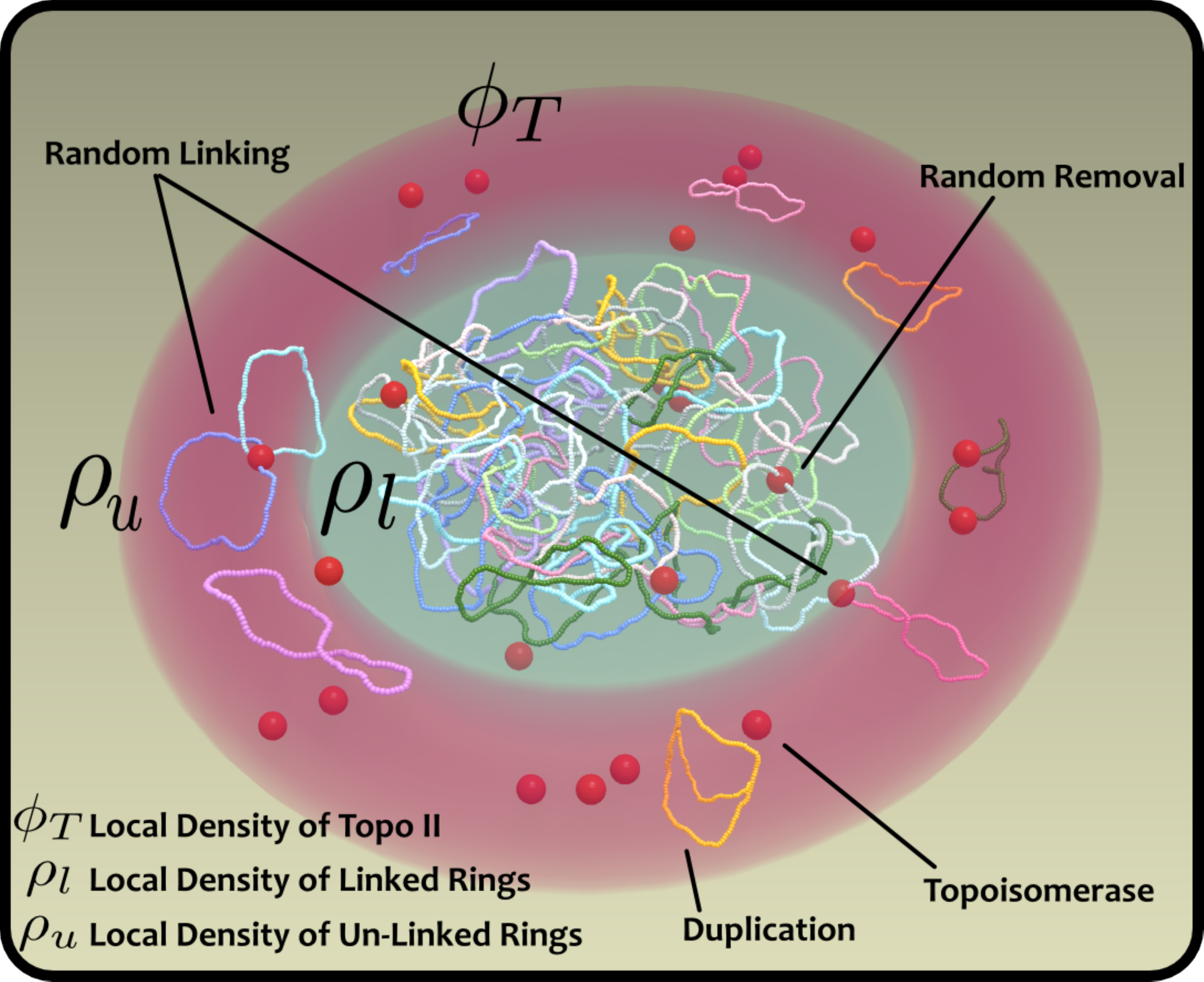}
\caption{Graphical representation of the full model. }
\end{figure*}

The other terms in eqs.~\eqref{eq:Eqs.a}-\eqref{eq:Eqs.c} are all justified in terms of the reactions described in eqs. \eqref{eq:removal}-\eqref{eq:reattachment}. The value of $\alpha$ can be estimated by noting that polymerase replicates DNA at a speed between $20$ $bp/s$ (in bacteria) and $500$ $bp/s$ (in eukaryotes)~\cite{Dignam1983,Wickiser2005,Schwartz2009}. We obtain a typical duplication time  of about $\alpha^{-1} \simeq 2000$ $bp/100$ $bp/s = 20$ $s$. In addition, we set the rate of production of Topoisomerase II $\kappa = \alpha = 0.05$ $s^{-1}$. This is because of we assume that this process involves the transcription of mini-circles, which requires a time comparable to their duplication. Finally, the decay time $\tau$ associated to Topoisomerase enzymes has been set to $\tau = 600$ $s$, being this its typical half-life in bacteria~\cite{Taniguchi2010}. Having already motivated the choices of the diffusion constants, the only free parameters we are left with are $R$ and $A$. Here, we mention that the typical time-scale for these two quantities is related to the action of Topoisomerase and therefore the rate at which these two processes occur can be estimated to be roughly $\tau^{-1} \sim 0.001$ $s^{-1}$. In the next section, we will describe further the relationship between the removal rate $R$ and the attachment rate $A$. \\

\subsection{System units}
We solve equations \eqref{eq:Eqs.a}-\eqref{eq:Eqs.c} numerically, by subdividing the system into discrete segments of size $\sigma$. The size of the Kinetoplast is $1$ $\mu m$ and we set $10 \sigma = 1$ $\mu m$. The unit time is left in seconds. This leads a rescaling of the diffusion coefficients to $D_u =0.001$  $\mu m^2/s = 0.1$ $\sigma^2 /s$ and consequently $D_l = D_T = D_u/5000 = 0.00002$ $\sigma^2/s$.

\section{Results}
The initial condition representing a non-replicating Kinetoplast is $\rho_l=1$ for $r<R_k$ and $\rho_u=0$ everywhere. As long as we do not introduce some Topoisomerase in the system, \emph{i.e.} $\phi_T=0$, equations \eqref{eq:Eqs.a}-\eqref{eq:Eqs.c} describe a non-replicating system, as $\rho_u$ and $\phi_T$ will remain zero throughout the time evolution. As soon as we introduce some Topoisomerase in the system, the replicating process begins and sustains itself until the Kinetoplast has been fully replicated. This is a feature that emerges naturally from our model and plays the natural role of a switch for the replication initiation. As soon as the organism needs to replicate, a small amount of Topoisomerase is all it is needed for the whole replication to take place; here we choose $\phi_T = 0.1$ at $t=0$.\\

\begin{figure*}
\centering
\includegraphics[scale=0.4]{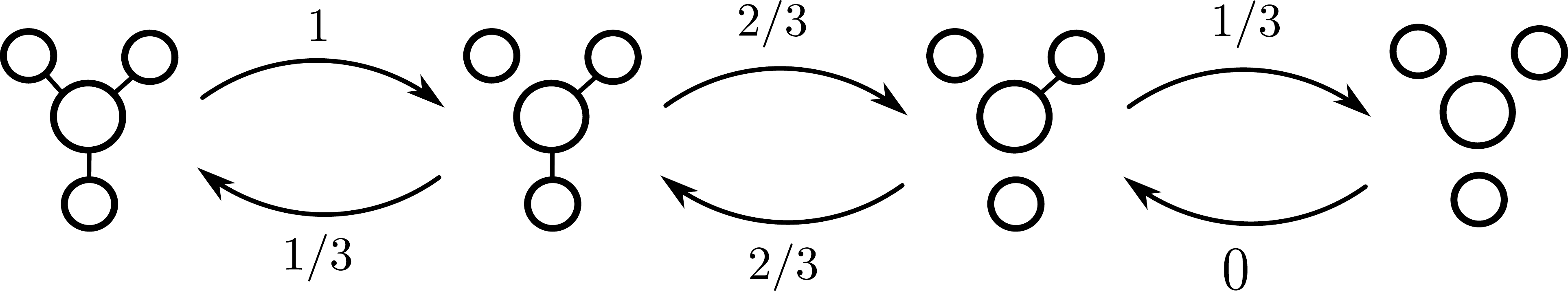}
\caption{Jumping probabilities between the possible linked states of single mini-circles within the Kinetoplast. Once the mini-circle in the middle has been freed, it immediately diffuses away. This makes the free state, an absorbing state for the mini-circles which have to be removed from the Kinetoplast.}
\label{fig:Rtermsnew}
\end{figure*}

\subsection{Rate of Removal $R$ and Rate of Attachment $A$}
\begin{figure*}[t]
\centering
\includegraphics[scale=0.35]{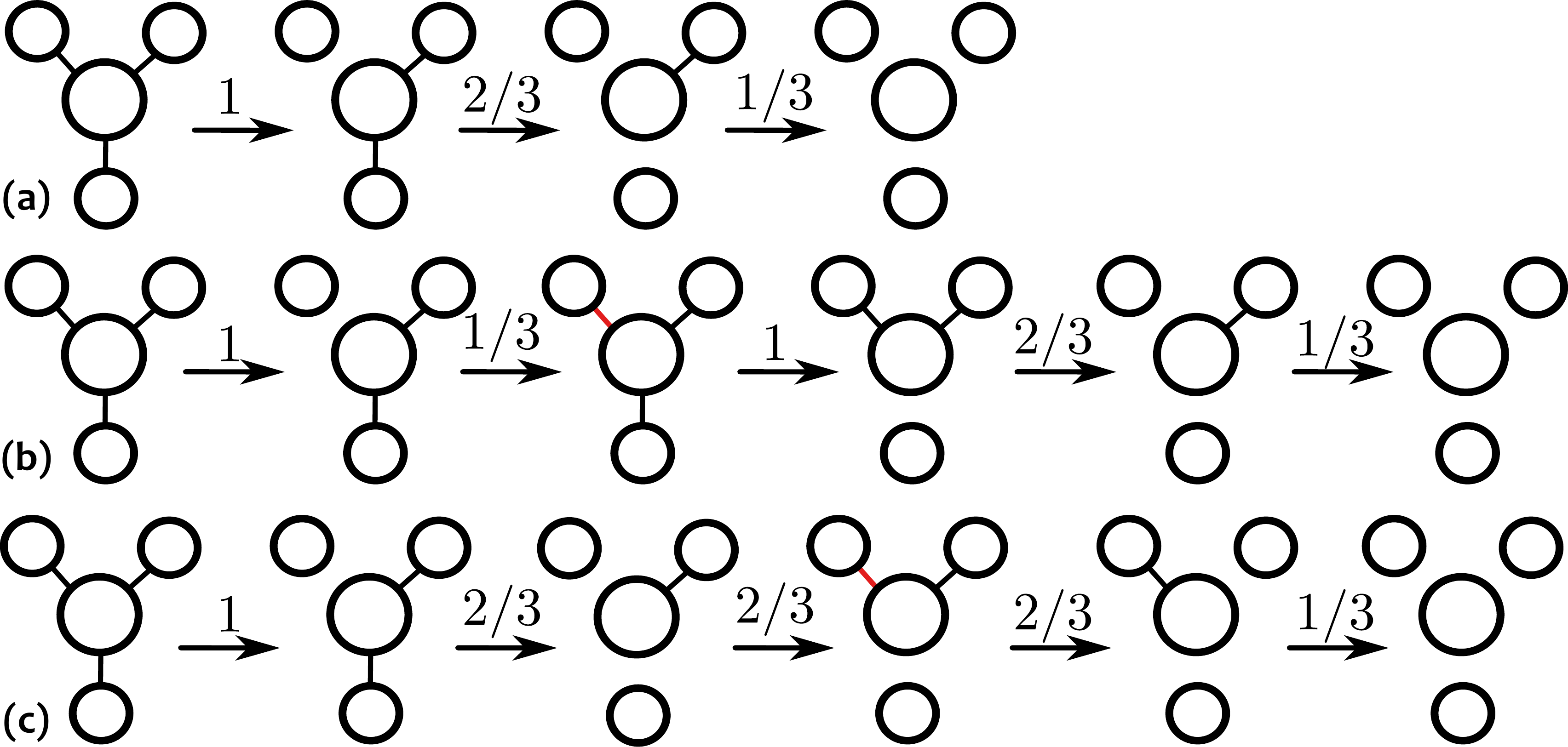}
\caption{Here we sketch three possible sequences for the complete freeing of mini-circles from the Kinetoplast, assuming that each mini-circle has initially three linked neighbours. The fractions on top of the arrows represent the probability associated with the action of Topoisomerase II. \textbf{(a)} Shows the most likely sequence, \emph{i.e.} the $n=0$ term, where no re-linking occurs.  \textbf{(b)-(c)} Show the $n=1$ terms, \emph{i.e.} the sequences where one new link is added during the process of freeing the middle mini-circle. In red the links which are newly created.}
\label{fig:Rterms}
\end{figure*}

The only free parameters in our model are $R$ and $A$, which  are, respectively, the rates of removal (or unlinking) of mini-circles from the network and of re-attachment (or linking) to either the network or other free mini-circles. The Kinetoplast network has been shown to be a percolating network at its critical point which implies that the number of linked neighbours per mini-circle is, on average, three~\cite{Chen1995,Diao2012a,Michieletto2014a}. This has also been confirmed by experiments to be the case for the Kinetoplast~\cite{Chen1995}. Because each mini-circle is topologically linked with, on average, three neighbours, we conjecture that the rate of removal of mini-circles from the network is smaller than the rate of re-attachment. This is justified by the fact that one strand crossing is enough to attach a mini-circle to the Kinetoplast, while more than one strand crossing are required to remove a mini-circle from the network. The quickest pathway to free a mini-circle needs three specific strand crossings, \emph{i.e.} those which unlink the mini-circle from its neighbours. Nevertheless, in general, the action of Topoisomerase can also link neighbouring rings to a mini-circle. We can therefore estimate the ratio between the rate of removal $R$ and the rate of reattachment $A$ by using the following reasoning. Each time a molecule of Topoisomerase attaches to a mini-circle, it can either unlink or link a neighbour to that mini-circle. By assuming that each mini-circle has only three neighbours, the probability of unlinking one of the three linked neighbours from a mini-circle is 1. Once that one neighbour has been unlinked, the next time a molecule of Topoisomerase attaches to the first mini-circle can either unlink one of the 2 remaining neighbours (with probability $2/3$) or re-attach the previously unlinked neighbour (with probability $1/3$), and so on, until the first mini-circle is freed from all the neighbours (see Fig.~\ref{fig:Rtermsnew}). The sequence in which all the neighbours are removed one-by-one without being re-attached has probability $p=2/9$ to happen (see Fig.~\ref{fig:Rterms}(a)). Since every strand crossing requires one action of Topoisomerase, we assume that it occurs at rate $A \simeq \tau^{-1}$, the same as an attachment event. This implies that the ratio between removal $R$ and re-attachment $A$ can be estimated via the pathways probability $p$. In fact, by taking into account the sequences in which Topoisomerase re-attaches back a neighbour $n$ times, we approximate the ratio $R/A$ as the probability of the average sequence $p_{\langle n \rangle}$. The distribution $p_n$ is computed as (see Fig.~\ref{fig:Rtermsnew})
\begin{equation}
p_n = \sum_{k=0}^n {n \choose k} \left[ 1 \times \left(\dfrac{1}{3} \times 1 \right)^k \times \dfrac{2}{3} \times \left(\dfrac{2}{3} \times \dfrac{2}{3}\right)^{n-k} \times \dfrac{1}{3}\right] =  \dfrac{2}{9} \left(\dfrac{7}{9}\right)^n . 
\end{equation}\\
The first moment of the distribution is $\langle n \rangle=3.5$, which gives, by interpolation, $p_{\langle n \rangle} = 0.092$. In light of this we use $R/A = 0.1 \simeq p_{\langle n \rangle}$ in what follows, acknowledging that this is merely a crude estimate.\\

\begin{figure*}[t]
\centering
\includegraphics[scale=0.66]{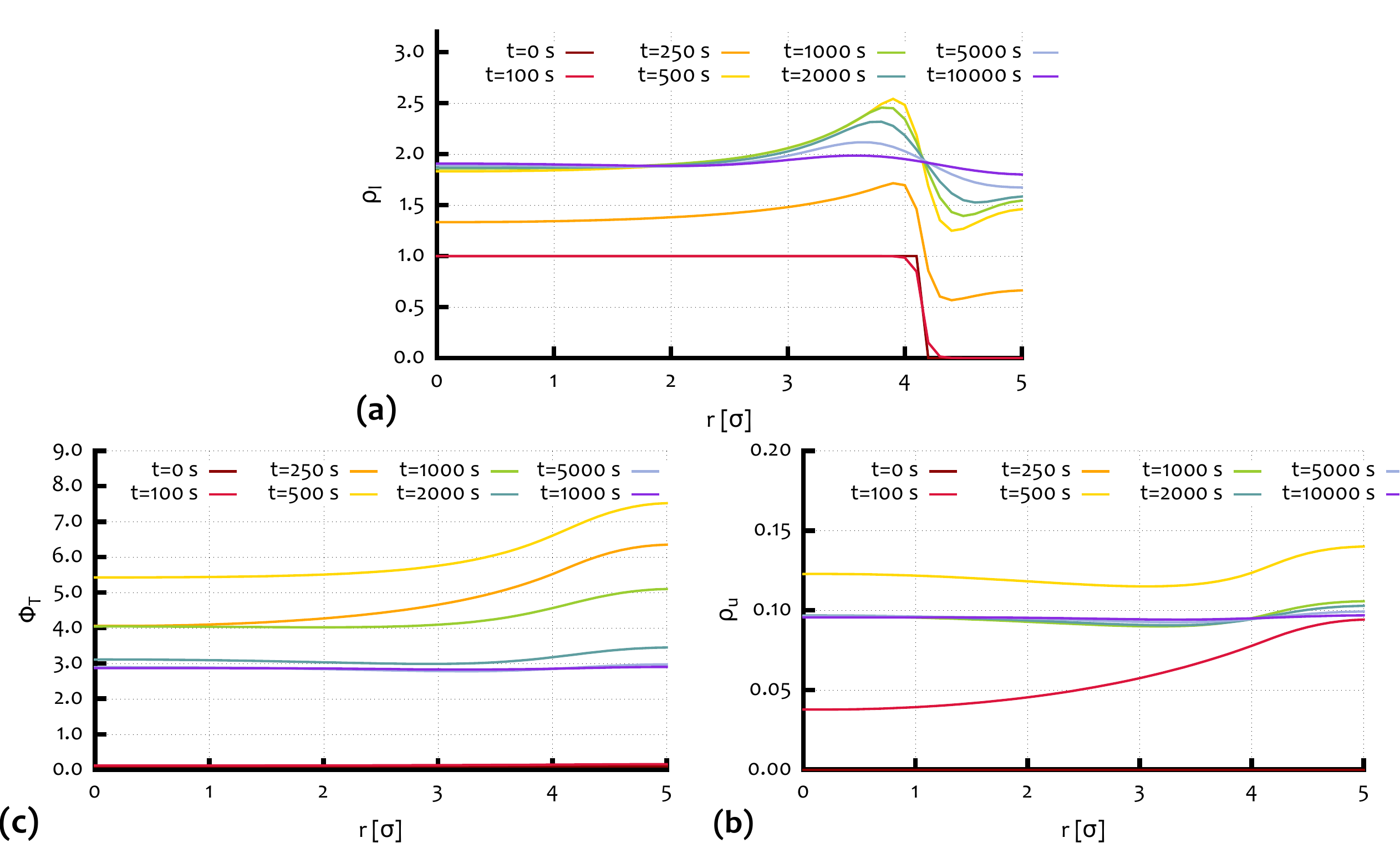}
\caption{\textbf{(a)} Profile of $\rho_l(r,t)$. Note the higher density of linked circles at the edge of the network, compatible with experimental findings~\cite{Jensen2012}. As time progresses, mini-circles are reattached to the Kinetoplast from the edge of the network, generating an apparent centripetal motion of progeny mini-circles toward the centre of the Kinetoplast. This can be seen here, as the peak of the distribution moves toward smaller values of $r$ as time progresses. Figures \textbf{(b)} and \textbf{(c)} show the concentration of un-linked mini-circles and Topoisomerase, respectively. Throughout the time evolution, the concentration of Topoisomerase remains higher at the edge of the network, once again in agreement with experimental findings~\cite{Shlomai1983,Shlomai1994}. }
\label{fig:Panel2}
\end{figure*}

The profiles of $\rho_l(r,t)$, $\rho_u(r,t)$ and $\phi_T(r,t)$ for this value of $R/A$ are shown in Fig.~\ref{fig:Panel2}. One can notice that free mini-circles show a higher concentration at the edge, where they have been observed to complete their duplication and to finally re-attach to the Kinetoplast. Correspondingly, one can notice a higher concentration of Topoisomerase in the same region, which is in quantitative agreement with experimental findings~\cite{Shlomai1983,Shlomai1994,Jensen2012}. Also, as time progresses, the distribution of linked mini-circles shows a peak at the edge of the network. This also shows an apparent centripetal motion toward the centre of the Kinetoplast, \emph{i.e.} the maximum of $\rho_l(r,t)$ shifts from high values of $r$ to small ones. This effect might be enhanced by the presence of a term describing steric repulsion in eqs.~\eqref{eq:Eqs.a} and \eqref{eq:Eqs.b}, \emph{e.g.} a contribution to the free energy density describing the system of the form $f_{steric} \sim \rho_l \rho_u$. We chose not to include such a term in order to keep our model as simple as possible, noting that it still allows us to reproduce relevant experimentally observed features.

\subsection{Fixed Point Analysis}

By studying the fixed points of the system of equations \eqref{eq:Eqs.a}-\eqref{eq:Eqs.c}, one can find that the fixed point value of $\rho_l$ is 
\begin{equation}
\rho_l^* = \dfrac{4}{R/A + 2},
\end{equation}
where we used the fact that, at equilibrium, $\rho_l$ and $\rho_u$ are related via $\rho_u + \rho_l =  2$. This means that, depending in the value of $R/A$, the newly formed Kinetoplast can be highly, but not fully, interlocked. Only in the case that $R/A$ was zero, then every mini-circle in the system would belong to the Kinetoplast network. Any state for which $R/A > 0$ corresponds to a \emph{marginally linked network}, \emph{i.e.} a percolating network which does not include all the mini-circles in the system. One can notice that, in order to obtain a \emph{fully linked network} of mini-circles at the end of the replication, \emph{i.e.} $\rho_l \simeq 2$, the Kinetoplast ought to send the ratio $R/A$ to zero. Amusingly, one of the ways of reducing this ratio is by creating very highly interlocked networks, \emph{i.e.} networks with high valence. In other words, in order to make highly linked networks, the system needs to start from highly linked networks. Because each mini-circle is linked to a high number of neighbouring rings, the removal rate would be drastically reduced.

\begin{figure}[t]
\centering
\hspace*{-1 cm}
\includegraphics[scale=0.7]{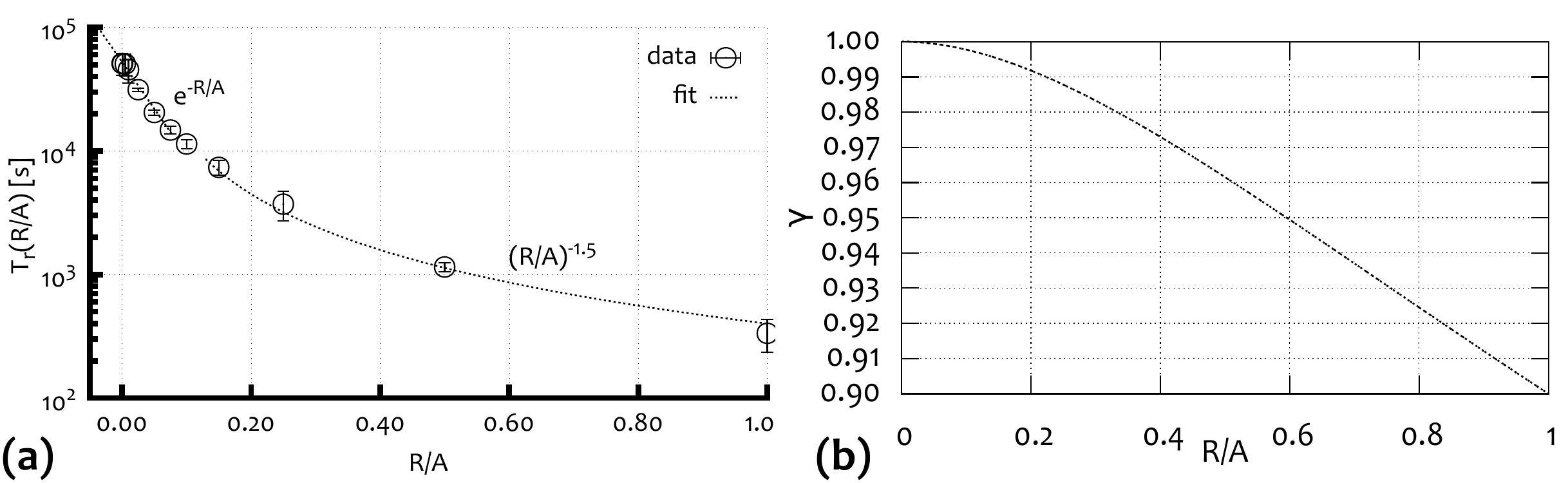}
\caption{\textbf{(a)} Behaviour of $T_r$ as a function of the ratio $R/A$. $T_r$ is defined as the time required for the function $\rho_l(r,t)$ to reach the steady state. In practice we require $\sum_r \rho_l(r,t+1) - \rho_l(r,t) < 0.001 $. By lowering the ratio $R/A$, the total replication time $T_r$ grows exponentially. A faster replication is obtained by increasing $R/A$.  \textbf{(b)} Solution of eq.~\eqref{eq:fitness} as a function of $R/A$. By decreasing $\gamma$, the fraction of unique genetic material contained in the mini-circles is lowered. This leads to higher redundancy and hence faster replication times (high $R/A$) since a larger number of mini-circles can be lost without putting the progeny fitness at risk.}
\label{fig:TEQ}
\end{figure}

On the other hand, by decreasing $R$, the time required to replicate the network in full ($T_r$) grows exponentially for small $R/A$, as we show in Fig.~\ref{fig:TEQ}. This implies that the Kinetoplast would need an exponentially long time to form a linked network with $\rho_l=2$ and $\rho_u=0$, and from an evolutionary point of view, this is not an advantage. It is therefore tempting to speculate that the value of $R/A$ has been tuned by evolution to balance the need for a well-linked network, \emph{i.e.} $\rho_l \simeq 2$, to that for a reasonably short time to complete replication.


The competition between these two effects can be summarised in one equation that describes the growth of a population of $N$ organisms as a function of $R/A$ and $\gamma$, which we call the critical lethal fraction of mini-circles unlinked from the Kinetoplast at the end of replication, beyond which the organisms are no longer viable. The equation reads:
\begin{equation}
\dfrac{1}{N}\dfrac{dN}{dt} = e^{R/A} \left[ 1 - \gamma \left(2 - \rho_l^*(R/A)\right)\right] = f(R/A).
\label{eq:fitness}
\end{equation}

According to eq.~\eqref{eq:fitness}, $N$ will grow at a rate proportional to the speed of replication, $i.e.$  proportionally to $T_r^{-1} \sim e^{R/A}$ and decay at a rate proportional to the fraction of mini-circles which are not linked to the network, \emph{i.e.} $\gamma \left(2 - \rho_l^*\right) =\gamma \rho_u^*$. The function $f$ measures the logarithmic growth rate of a population of Kinetoplasts and hence gives a measure of its evolutionary fitness. Maximising such fitness means setting $df(R/A)/d(R/A)=0$, which leads to the following relation:
\begin{equation}
\gamma(R/A) = \dfrac{\dfrac{1}{2}\left(\dfrac{R}{A}\right)^2 + 2\left(\dfrac{R}{A}\right) + 2}{\left(\dfrac{R}{A}\right)^2 + 2\left(\dfrac{R}{A}\right) + 2}.
\label{eq:fitness_min}
\end{equation}

This curve (shown in Fig.~\ref{fig:TEQ}) gives a relation between the ratio $R/A$ and the parameter $\gamma$. As mentioned before, the latter describes the critical lethal fraction of mini-circles unlinked from the Kinetoplast at the end of the replication, \emph{i.e.} $\rho_u^* = 2 - \rho_l^*$. Because each mini-circle carries genetic material, the parameter $\gamma$ can also be interpreted as ``redundancy'' in the genetic material carried by the mini-circles. In fact, by setting $\gamma=1$, we need to have $R/A=0$ in order to satisfy eq.~\ref{eq:fitness_min}. This implies that every mini-circle needs to be linked to the network at the end of the replication, \emph{i.e.} every mini-circle is crucial for the survival of the organism (see eq. \eqref{eq:fitness}). This can be interpreted with the fact that, in this case, each mini-circle carries unique genetic information, and any loss of mini-circles would cause the reduction of evolutionary fitness~\footnote{Here, we are assuming that any mini-circle which is not linked to the Kinetoplast at the end of the replication is lost during the separation into progeny cells.}. In other words, setting the  parameter $\gamma =1$, signifies that there is no redundancy in the genetic material carried by the mini-circles, and each one of them has to be passed on to the progeny in order for the organism to survive. On the other hand, it has been seen both experimentally and previously in this work, that the time required to fully replicate the network is on the time-scale of the one or two hours~\cite{Jensen2012}. This can be achieved by setting $0 < R/A \lesssim 0.3$ (from Fig.~\ref{fig:TEQ}(a)), which implies that the parameter $\gamma$ has to be strictly smaller than unity (from eq.~\ref{eq:fitness_min} and Fig.~\ref{fig:TEQ}(b)). In other words, the mini-circles carry heterogeneous, but non unique, genetic material. This is once again in agreement with experimental findings, which have observed that although the mini-circles are heterogeneous in the genetic material, they also show, to some extent, redundancy of the genetic information~\cite{Jensen2012}. It is also worth noting that for a broad range of values for $R/A$, the value of $\gamma$ stays close to $1$,  meaning that the system does not need a large redundancy in the genetic material to ensure survival, which is again qualitatively consistent with experimental observations, which showed that most of the mini-circles are genetically heterogeneous.

\section{Conclusions}
Our model naturally connects the topological state of the Kinetoplast DNA with the genetic information encoded in the mini-circles. We propose an analytical model for the replication of the Kinetoplast DNA. This is treated as a two dimensional disk, surrounded by a shell which plays the role of the anti-podal regions where the enzymes necessary for the duplication and transcription of the mini-circles are enriched. We assume that the genetic material of the mini-cirles contains transcription factors which promote the production of the Topoisomerase enzymes. We find that the replication mechanism can be described as a self-regulated process, requiring only the input of a small initial level of Topoisomerase to fully replicate the Kinetoplast. We find that the ratio between rate of removal $R$ and rate of attachment $A$ of mini-circles from and to the Kinetoplast controls how the linked network is contracted. Our model qualitatively reproduces the experimentally observed higher concentration of Topoisomerase enzyme at the edge of the network and the apparent centripetal motion of newly linked mini-circles from the edge of the network toward its centre. It also reproduces the typical duplication time-scale, which has been found to be of the order of  hours. The final topological state of the network is found to be  controlled by the ratio $R/A$ with the degree of linkage decreasing with increasing $R/A$. On the other hand, we find that lowering this quantity leads to exponentially longer replication times, which decrease the evolutionary fitness. By considering these two effects, we propose an explicit formula for the growth of a population of these organisms, which gives a relationship between accuracy and speed of the replication mechanism as a function of the ``redundancy factor'' $\gamma$. This factor can be interpreted either as the lethal fraction of mini-circles unlinked from the network at the end of replication or as a measure of the degree of uniqueness (absence of redundancy) of crucial genetic material carried by the mini-circles. 

\section*{ACKNOWLEDGEMENTS}
DMi acknowledges the support from the Complexity Science Doctoral Training Centre at the University of Warwick with funding provided by the EPSRC (EP/E501311). We also acknowledge the support of EPSRC to DMa, EP/I034661/1, and MST, EP/1005439/1, the latter funding a Leadership Fellowship.

\bibliographystyle{iopart-num}
\bibliography{KinetoplastReplicationFIXEDnourl}

\end{document}